\documentclass[12pt]{elsarticle}
\usepackage{amsmath,amssymb}
\usepackage[utf8]{inputenc}
\usepackage{xcolor}
\usepackage{subcaption}
\usepackage{graphicx}

\begin{document}

\begin{frontmatter}

\title{Fractal Social Dynamics as a Driver of Consensus and Inequality}
\author{Airton Deppman\corref{cor1}}
\ead{deppman@usp.br}
\cortext[cor1]{Corresponding author}
\address{Institute of Physics \\ University of São Paulo \\ São Paulo, Brazil}

\begin{abstract}
Human social behavior is organized in stratified, hierarchical networks, with a support group with about 5 members, expanding proportionally at each layer up to a maximum of approximately 150 frequent interactions per individual. This is known as Social Brain Hypothesis, and its findings are supported by psychological and neurological evidence. The fractal network framework provides valuable insights into social phenomena such as the spread of fake news and the development of technology. This study models socioeconomic interactions using fractal networks, where group sizes scale by a fixed factor, to analyze how consensus is formed. Using $q$-calculus, the model reveals how hierarchical structures influence information spread, highlighting universal features governed by power laws.. The results follow $q$-Gaussian distributions, showing heavy-tails that align with observed inequalities in societies worldwide. The results show that inequalities arise from the fractal structure of the socioeconomic network.
\end{abstract}


\end{frontmatter}

\section{Introduction}

There is strong evidence that human social behavior, as well as that of other primates, is structured in layered social systems. Individuals maintain a small number of close relationships, while also belonging to increasingly larger circles of social connections. This stratified organization has been observed across many cultures and is thought to arise from cognitive and emotional constraints, according to the Social Brain Hypothesis \cite{Dunbar1998}. One key finding is that the number of frequent, stable relationships a human can maintain is approximately 150, a value known as Dunbar’s number \cite{Dunbar1992}.

Within this structure, the closest social tier, the support group, typically includes only four to six individuals. Successive layers expand by roughly constant factors, giving rise to a hierarchical organization of social groups \cite{Zhou2005}. This layered scaling suggests that human societies may be effectively modeled as hierarchical networks, where each layer represents a level of social closeness. When these layers follow a regular scaling pattern, the resulting structure is called a fractal network. In this framework, individuals occupy the lowest layer, forming tightly connected groups. Each group acts as a unit in the next layer, which is composed of groups of groups, and so on. The size of each successive layer decreases by a fixed factor, typically related to the number of individuals in a support group.

Networks structured by hierarchical levels that scale by a fixed factor are known as fractal networks \cite{Song2006}. In this type of architecture, individuals form the base layer of the network. Each person connects to a group of $N_c$ closest contacts, forming a support group. Each support group then acts as a single agent in the next hierarchical layer. The number of agents in this second layer is $N/N_c$, where $N$ is the total number of individuals in the network. Each agent in this layer connects to $N_c$ others, and the process continues through subsequent layers, with the number of agents decreasing by a factor of $1/N_c$ at each level. This model idealizes real social organization, but in reality, agents at each layer overlap due to interconnected social ties, the scaling factor varies slightly, and the boundaries between layers become increasingly indistinct. Despite these limitations, fractal networks provide a valuable framework for investigating scale invant aspects of social structure~\cite{Deppman2021}. While they can only offer general insights into social dynamics, they are particularly effective at highlighting universal features that can be studied using the rigorous methods of the natural sciences.

Fractal networks of this kind have been applied to model a wide range of phenomena, including the spread of information, the dynamics of epidemics \cite{Newman2002}, and the formation of prices in financial markets~\cite{Stanley2001}. In the study of urban life, they help explain the emergence of universal patterns governed by simple mathematical functions known as power-laws \cite{Bettencourt2007}. For example, the so-called fundamental allometry reveals a nonlinear relationship between population size and urban area, which increases sublinearly \cite{Batty2008}. This is one of many cases suggesting that cities may be more efficient and less chaotic than previously believed.

These findings support the hypothesis that human social organization follows a hierarchical structure rooted in small, close-contact groups, typically composed of about five individuals. Psychological and neuroscientific evidence, including the Big Five personality model and fMRI studies \cite{McCrae1992}, suggests that cognitive constraints drive the formation of small, stable groups, providing a biological basis for modeling society as a fractal network. In contrast to Ref.~\cite{Arvidsson2023}, which emphasizes inequalities as a primary driver of urban scaling, this work posits that fractal social structure shepe the urban areas, with inequalities arising from the hierarchical interactions~\cite{Deppman2025, Batty2008}.

\section{Theoretical Background}

Fractal networks provide a powerful framework for modeling complex systems. These networks are characterized by self-similar, hierarchical structures, where agents are organized into layers that scale by a fixed factor \cite{Song2006}. Within this architecture, information typically spreads locally—from one individual to a small, fixed number of connected peers. This local transmission drives a wide range of phenomena, including social communication, epidemic outbreaks, and opinion dynamics \cite{Deppman2021}.

The spread of information in such networks is governed by a \textit{q}-exponential function, which emerges naturally from Tsallis statistics \cite{Tsallis2009}. This statistical framework captures the \textit{nonextensive} nature of interactions in complex systems, where correlations and memory effects are significant. At the mathematical core of this formulation lies the \textit{q}-derivative, which, through recent developments, has been shown to arise as a Continuous Approximation to the fractal derivative, describing how quantities evolve in fractal geometries \cite{Deppman2023}. This equation is central to the \textit{q}-deformed calculus framework \cite{Borges2004}, originally introduced to formalize the algebraic and differential structure underlying Tsallis statistics. While q-calculus was not initially linked to fractal geometry, later work demonstrated that it can emerge naturally from fractal calculus when continuous approximations are applied to functions defined on fractal supports \cite{Deppman2023}. This result establishes a deeper mathematical and physical connection between nonextensive statistics and the geometric complexity of fractal systems, and it plays a central role in modeling the dynamics explored in the present work.

In parallel, continuous opinion models—such as those incorporating bounded confidence—offer complementary insights into consensus formation. In these models, agents adjust their opinions only when differences fall below a certain threshold, reflecting a realistic cognitive bias in social interactions \cite{Hegselmann2002}.

To quantify the spread of information over time, let $i(t)$ represent the number of informed individuals at time $t$. A general expression for this function, within the fractal social model, is given by~\cite{Deppman2021}
\begin{equation}
i(t) = \left[1 + (1 - q) \frac{\kappa u(t) t}{\lambda} \right]^{\frac{1}{1 - q}},
\end{equation}
where $\kappa$ is the information transmission rate, $u(t)$ is the number of uninformed individuals at time $t$, $1 - q = 1/N_c$, with $N_c$ denoting the number of close contact among agents in the layer, and $\lambda$ is a scaling factor. At the individual level, $\lambda=1$. To maintain consistency with the mathematical formalism in Ref.~\cite{Deppman2025}, we apply a transformation $q \rightarrow 2 - q$, known as the \textit{parametric reflection} of the \textit{q}-exponential \cite{Deppman2021}. This preserves the interpretative meaning of $q$ while facilitating the analytical treatment.

This formulation captures the dynamics of information diffusion in a stratified network, showing a $q$-exponential increase in $i(t)$ over time, a characteristic feature of systems with non-linear and non-Markovian behaviors. The simplicity of this model rests on key assumptions. It treats information as a binary variable: individuals are either informed or uninformed, with no intermediate states. Furthermore, once informed, individuals remain so permanently, implying a one-way, irreversible process. While this is suitable for modeling basic information spread, many aspects of socioeconomic activity are more nuanced. For example, variables like wealth, influence, job opportunities, or technological development tend to evolve gradually over time and can increase or decrease. To model such quantities, a more general formulation is needed.

This generalization introduces a continuous variable $x$, which can assume both positive and negative values. A part $\kappa u(x,t)$ of the population can trade a quantity $x$ of the output at time $t$. The generalized function $f(x, t)$, describing the distribution of a variable over individuals and time, is then given by
\begin{equation}
f(x, t) = \left[1 + (q - 1) \frac{\kappa u(x, t)}{\lambda} \right]^{\frac{1}{q - 1}}.
\end{equation}
This expression recovers the previous binary case when $u(x, t) = \delta(x)u(t)t$, where $\delta(x)$ is the Dirac delta function. In this context, $f(x,t)$ gives the distribution of the output $x$ across the population at any time $t$.

Under this assumption, differentiating $f(x, t)$ with respect to $x$ yields:
\begin{equation}
\frac{\partial f}{\partial x} = \frac{\kappa}{\lambda} \left[1 + (1 - q)\frac{\kappa u(x, t)}{\lambda} \right]^{\frac{1}{1 - q}} \frac{\partial u}{\partial x}.
\end{equation}
Assuming $u(x, t)$ is a linear function in $x$ and $t$, we can take $\partial u / \partial x = k_x$ as constant, leading to:
\begin{equation}
\frac{\partial f}{\partial x} = \frac{k_x \kappa}{\lambda} f^q.
\end{equation}

This equation is central in the \textit{q}-deformed calculus framework \cite{Borges2004}, developed to describe non-additive processes in Tsallis statistics. It can be interpreted as a continuous approximation of the fractal derivative \cite{Deppman2023}, and is also related to correlation functionals in the Boltzmann equation under strong correlations \cite{Deppman2023b}.

In the context of this work, the stochastic process associated with consensus formation can be modeled by a generalized Boltzmann equation:
\begin{equation}
\frac{df}{dt}(x, t) = \int dy \, dz \, \left| M(y,z;x,t) \right|h[f(z), f(y)], \label{eq:transition}
\end{equation}
where the correlation functional is defined as
\begin{equation}
h[f(z), f(y)]_q = f(z) \otimes_q f(x) - f(y) \otimes_q f(x),
\end{equation}
with the q-product defined as \cite{Tsallis2009}:
\begin{equation}
f(y) \otimes_q f(z) = \left[ f(y)^{1 - q} + f(z)^{1 - q} - 1 \right]^{\frac{1}{1 - q}}.
\end{equation}
In the equations above, $z - y = 2x$ representing the transition probabilities in the system. For the limiting case $x \rightarrow 0$, we have $z \rightarrow y$, simplifying the functional. Un the equation above, $M(y,z;x,t)$ incorporates the microscopic  mechanisms regarding the transitions from the states $z$ to $x$ and $x$ to $y$. As $q \rightarrow 1$, the $q$-product becomes identical to the standard product~\cite{Borges2004}.

A key result from Ref.~\cite{Deppman2023b} shows that in the limit $x \rightarrow 0$, this expression reduces to the q-derivative above. Thus, the Boltzmann equation simplifies to
\begin{equation}
\frac{df}{dt} = C(x, t)  \frac{k_x \kappa}{\lambda} f^q \,,
\end{equation}
where $C(x,t)$ depends o the matrix elements in Eq.~(\ref{eq:transition}). This subtle shift—from the standard product to the q-product—captures the transition from extensive to nonextensive dynamics. The resulting formalism accounts for strong correlations, memory effects, and the underlying fractal geometry of the system.

Following a standard procedure, it can be shown that the generalized Boltzmann equation leads to the Plastino-Plastino Equation (PPE) \cite{Deppman2023b}, a non-linear generalization of the Fokker-Planck Equation (FPE), expressed as
\begin{equation}
\frac{\partial f}{\partial t} = -\frac{\partial}{\partial x}[A(x)f] + \frac{\partial^2}{\partial x^2} \left[ D(x) f^{2 - q} \right],
\end{equation}
where $A(x)$ represents a drift term (such as social pressure), and $D(x)$ is a diffusion coefficient. When $q \rightarrow 1$, the standard FPE is recovered. In principle, $A(x)$ and $D(x)$ can be calculated by the transition term in Eq.~(\ref{eq:transition}), in a model that describes the trasnition matrix $M$.

A simplified version of the equation assumes no drift $A(x) = 0$ and constant diffusion $D(x) = D$, yielding:
\begin{equation}
\frac{\partial f}{\partial t} = -\gamma \frac{\partial f}{\partial x} + D \frac{\partial^2 f^{2 - q}}{\partial x^2}.
\end{equation}
The stationary solutions of this equation are q-Gaussians, given by
\begin{equation}
f(x,t) = \frac{A}{\sigma(t)} \left[1 - (1 - q)\frac{(x + \gamma t)^2}{\sigma(t)^2} \right]^{\frac{1}{1 - q}},
\end{equation}
which describe peaked distributions of opinions or other social quantities. In the formula above, $x$ represents the fraction of wealth (or any other output) in the hands of a sector of the total population, and $\\gamma$ is the rate of change of that output \textit{per capita}. The parameter $q$ controls the tail behavior, with heavier tails corresponding to greater opinion diversity and weaker consensus. The function $\sigma(t)$ is limked to the distribution width, and is given by
\begin{equation}
 \sigma(t)^2=\left(\frac{t}{\tau}\right)^{\alpha}\,, \label{eq:sigma}
\end{equation}
where 
\begin{equation}
 \tau=\frac{3-q}{2-q} \frac{A^{q-1}}{2 D}\,,
\end{equation}
is a charactristic relaxation time, and $\alpha^{-1}=3-q$~\cite{Tsallis1996}.

\begin{figure}[ht]
\centering
\begin{subfigure}{0.49\textwidth}
\centering
\includegraphics[width=1\linewidth]{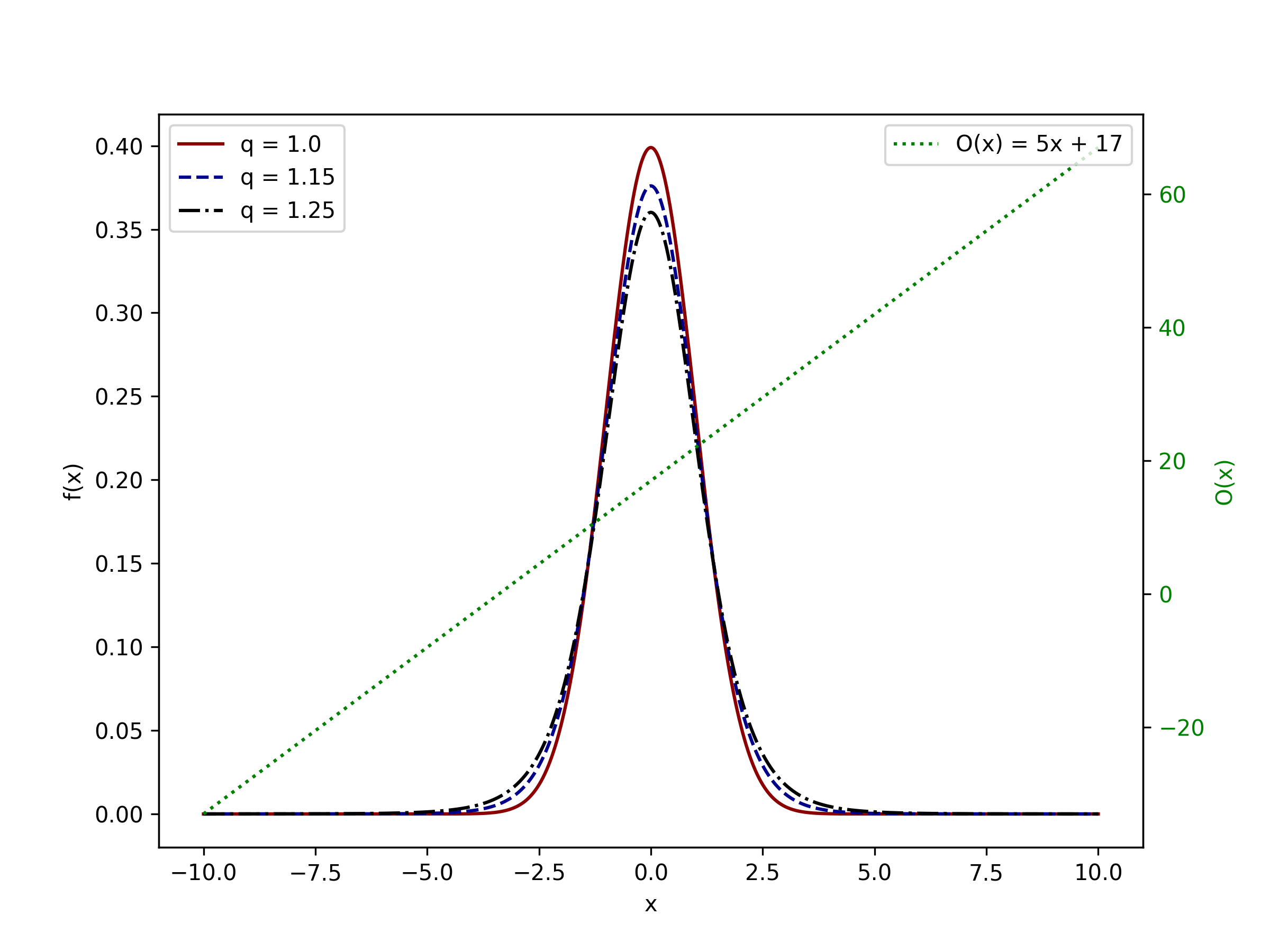}
\caption{q-Gaussian distributions for $q=1.0$, $q=1.15$ and $q=1.25$ (left axis). The function $O(x)=5x+17$ used as an application is shown (right axis).}
\label{fig:qgaussian}
\end{subfigure}
\begin{subfigure}{0.49\textwidth}
\centering
\includegraphics[width=1\linewidth]{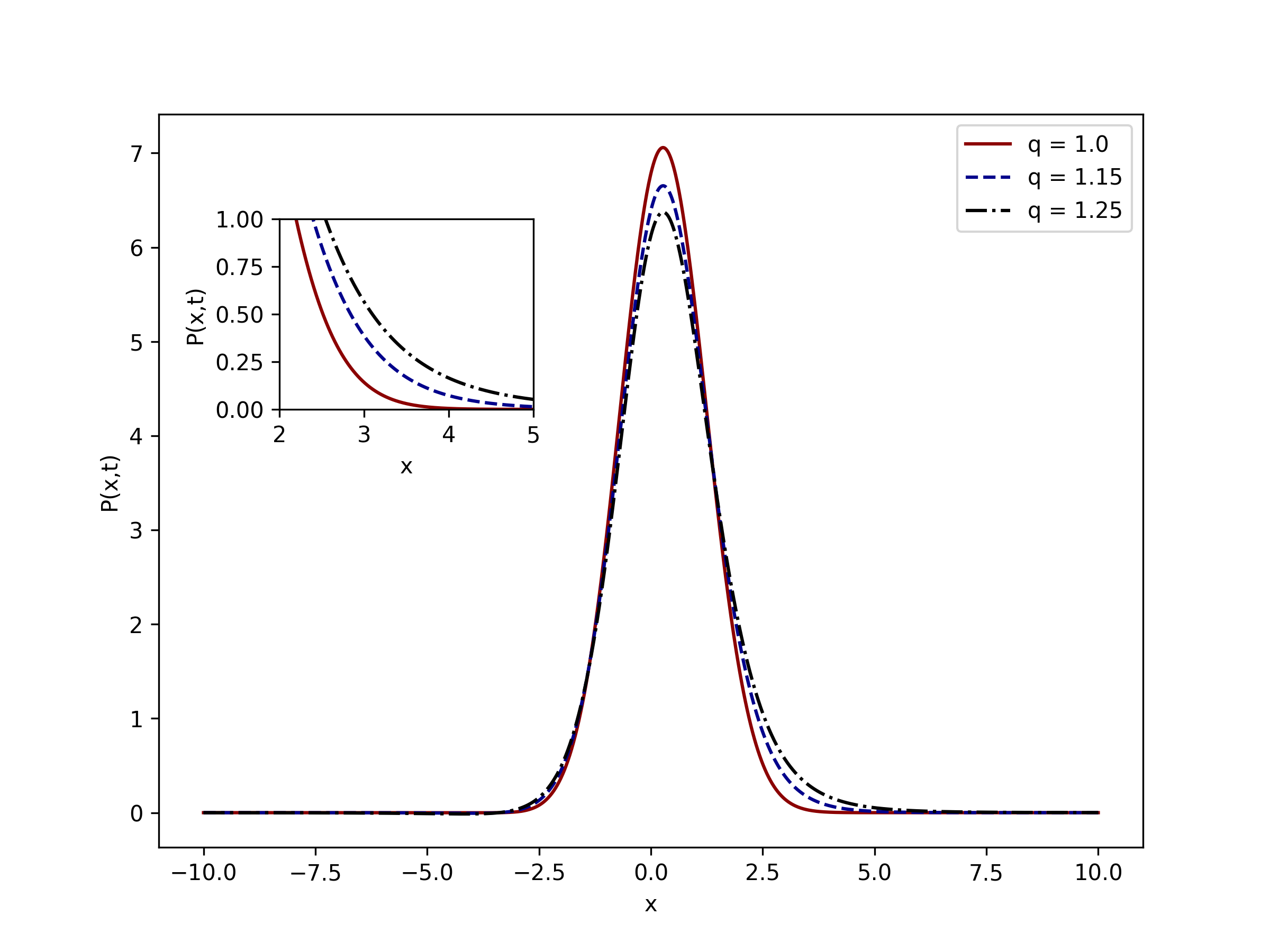}
\caption{Product $P(x,t)$ for different $q$ values, with an inset showing a zoomed-in view, highlighting the asymmetrical effects of the social ouput production.}
\label{fig:product}
\end{subfigure}
\caption{Visual representation of q-Gaussian distributions and their product with a linear function, illustrating the effect of varying $q$ on distribution shapes.}
\label{fig:results_plots}
\end{figure}

The difference between the Gaussian and the $q$-Gaussian distributions appears more clearly at the tails: the larger $q$ is with respect to 1, the heavier the distribution tails are. This can be observed in Fig.~\ref{fig:qgaussian}. When the parameter $q=1$, the Gaussian is recovered, and the plots show that in this case the tails are small compared with the other cases. This means that the larger the value for $q$, the more unequal the distribution of the quantity $x$ is in the society considered here. From the relation between $q$ and $N_c$, it means that the smaller the size of the support group is, the larger the inequalities are.

Some aspects of the result obtained here were already discussed in relation to the superlinear behaviour of social output. In Ref.~\cite{Arvidsson2023} it was shown that the superlinear behaviour disappears if heavy tails are cut. This can be easily understood in light of the present results. Assuming a linear output of the quantity $x$, that is, $O(x)=Ax+B$ for the entire population, that is, the output $O(x)$ of individuals is proportional to the quantity of $x$ they already retain, the distribution of output throughout the society is obtained by the product $P(x,t)=f(x,t) O(x)$. To illustrate the result, the case $O(x)=5x+17$ is considered, and the resulting distribution of the output distribution is plotted in Fig.\ref{fig:product}, where it can be observed that the right-hand tail contribution is larger than the left-hand contribution.

Observe that, in comparison with the Gaussian distribution ($q=1$), the $q$-Gaussians present higher contributions to the output, and the higher $q-1$ is, the larger the output is for the same value of $x$, as can be clearly observed in the inset of Fig.~\ref{fig:product}. The cumulative advantage mechanism proposed in Ref.~\cite{Arvidsson2023} aligns with our model, as hierarchical structures could enhance opportunities for top performers, leading to heavier tails in larger systems. Cutting the tails approximates the output distributions to the Gaussian case, as can be seen by inspecting the inset, so the result tends to be an output that is linear with the population size, as happens with Gaussian distributions.

Ref.~\cite{Arvidsson2023} correctly shows that superlinear urban scaling arises from heavy-tailed distributions, which reflect within-city inequalities. The present model reproduces this effect using q-Gaussians (Fig.~\ref{fig:product}), confirming that tail effects drive superlinear output. However, this work proposes that the fractal structure of social networks, rather than inequality itself, is the primary cause, with heavy tails emerging as a natural consequence of hierarchical interactions.

However, if inequalities are a natural consequence of our social structure, which is hardwired in our brains, why inequalities result so diffenrent among countries and regions worldwide? The answer relies on the dynamics of the output distribution. This work presents a simple example fo a linear output production, shown in Fig.~\ref{fig:product}, which illustrates how the differences in output production across the society distorts the original $q$-Gaussian distribution, by rising the right tail more than the left tail. The dynamics determines how such a distortion is corrected along time, by reaching a new equilibrated distribution with new peak position and width, and the form how the new distirbution is reached depends on the transport coefficients. The source of pronounced and persistent inequalities can be attributed to differences in the transport coefficients.

Therefore, the theoretical approach presented here can be used as as a framework for developing microscopic models, by stuying the effects of the transition matrix $M$ on the transport coefficients. Analyses of this kind can contribute to public policies that will favor less inequal societies by providing clues on the more effective pratices that allow a faster an better social output distribution.

\section{Conclusion}

This study establishes a fractal network framework for consensus formation, using $q$-calculus to model nonextensive dynamics. The Plastino-Plastino Equation (PPE), derived here from a generalized Boltzmann Equation, offers insights into diverse social and economic processes, including fake news propagation, price formation, and technological development. These systems exhibit power-law behaviors as universal signatures, rooted in the underlying hierarchical structure of social networks. The modeling is based on findings from the Social Brain Hypothesis and from the Big Five personality traits.

The results demonstrate that consensus formation and socioeconomic inequality emerge naturally from the fractal organization of human social groups. By linking local dynamics—governed by close-contact interactions—to global-scale phenomena, the model shows that opinion polarization and urban output scaling are manifestations of the network's self-similar geometry and the nonextensive nature of information flow.

In contrast to models emphasizing inequality as a primary driver \cite{Arvidsson2023}, this approach reveals that inequality, observed in urban productivity and wealth distributions, stems from the combination of fractal social architecture and linear individual outputs. The tails of the $q$-Gaussian distributions significantly contribute to superlinear scaling, consistent with recent empirical findings \cite{Arvidsson2023}. Future research should validate these predictions using empirical data from urban metrics, financial markets, or social media platforms, testing for $q$-Gaussian distributions and exploring more realistic dynamics by relaxing assumptions such as constant uninformed population or linear diffusion. For instance, testing q-Gaussian distributions in wealth data from cities like São Paulo or New York could confirm the model’s predictions.

\bibliographystyle{ieeetr}
\bibliography{SocialConsensus.bib}

\end{document}